\documentclass[aps,prl, notitlepage,reprint,superscriptaddress]{revtex4-2}

\usepackage{changes}

\usepackage[colorlinks,linkcolor=red,citecolor=blue,urlcolor=red]{hyperref}

\usepackage{graphicx}%,float} 
\usepackage{ulem}
\usepackage{amsmath,amssymb,amsfonts} % standard AMS packages
\usepackage{bm}	% bold symbols in math mode \bm{...}
\renewcommand{\mathbf}{\bm}

\usepackage{dsfont}	% proper mathbb format
\renewcommand{\mathbb}{\mathds}	% redefine \mathbb

\usepackage{mathrsfs} % use \mathscr{} for script letters in math
\usepackage{mathtools} % for proper typesetting of := and =:

\usepackage{color}
\usepackage{siunitx}
\usepackage{multirow}

\newcommand{\fref}[1]{Fig.~\ref{#1}}
\renewcommand{\eqref}[1]{Eq.~\ref{#1}}

\begin{document}
	
	\title{Microwave-to-optical conversion with a gallium phosphide photonic crystal cavity}

	\author{Simon H\"{o}nl}
	%\author{\color{red}J. Pinkman}
	\affiliation{IBM Research Europe, Zurich, S\"{a}umerstrasse 4, CH-8803 R\"{u}schlikon, Switzerland}
	
	\author{Youri Popoff}
	\affiliation{IBM Research Europe, Zurich, S\"{a}umerstrasse 4, CH-8803 R\"{u}schlikon, Switzerland}
	\affiliation{Integrated Systems Laboratory, Swiss Federal Institute of Technology Zurich (ETH Z\"{u}rich), CH-8092 Z\"{u}rich, Switzerland}
	
	\author{Daniele Caimi}
	\affiliation{IBM Research Europe, Zurich, S\"{a}umerstrasse 4, CH-8803 R\"{u}schlikon, Switzerland}

	\author{Alberto Beccari}
	%\author{\color{red}T. Alquist}
	\affiliation{Institute of Physics, Swiss Federal Institute of Technology Lausanne (EPFL), CH-1015
		Lausanne, Switzerland}
	
	\author{Tobias J. Kippenberg}
	%\author{\color{red}T. Alquist}
	\affiliation{Institute of Physics, Swiss Federal Institute of Technology Lausanne (EPFL), CH-1015
		Lausanne, Switzerland}

	%\author{\color{red}W. White}
	\author{Paul Seidler}
	\email{pfs@zurich.ibm.com}
	\affiliation{IBM Research Europe, Zurich, S\"{a}umerstrasse 4, CH-8803 R\"{u}schlikon, Switzerland}
	
	\date{\today}
	
	\begin{abstract}
	%Microwave-to-optical conversion is a prominent application for optomechanical systems and could enable future optical quantum networks between quantum computing platforms based on qubits at microwave frequencies. 
	Electrically actuated optomechanical resonators provide a route to quantum-coherent, bidirectional conversion of  microwave and optical photons. Such devices could enable optical interconnection of quantum computers based on qubits operating at microwave frequencies.
	Here we present a novel platform for microwave-to-optical conversion comprising a photonic crystal cavity made of single-crystal, piezoelectric gallium phosphide integrated on pre-fabricated niobium circuits on an intrinsic silicon substrate. 
	The devices exploit spatially extended, sideband-resolved mechanical breathing modes at ${\sim\SI{3.2}{GHz}}$, with vacuum optomechanical coupling rates of up to $g_0/2\pi \approx\SI{300}{kHz}$.  
	The mechanical modes are driven by
	%Coupling of the photonic crystal cavity to 
	integrated microwave electrodes 
	%occurs 
	via the inverse piezoelectric effect. 
	We estimate that the system could achieve an electromechanical coupling rate to a superconducting transmon qubit of $\sim\SI{200}{kHz}$.  
	%We consider this a 
	Our work represents a decisive step towards integration of piezoelectro-optomechanical  interfaces with superconducting quantum processors. 
	
	\end{abstract}

\maketitle
\section*{Introduction}
%The advent of quantum computers based on superconducting circuits and their recent rapid development \cite{arute2019quantum,jurcevic2021demonstration} have led to the pursuit of a plethora of strategies for scaling to larger numbers of qubits to create sytems that %vastly 
%outperform classical computers.
Recent years have seen the advent of quantum computers based on superconducting microwave circuits and their  rapid development \cite{arute2019quantum,jurcevic2021demonstration} towards commercial systems that outperform classical computers.
Due to their low energy, the microwave qubits are bound to the %ultracold 
millikelvin environment of  a dilution refrigerator, which presents a variety of challenges for building systems with large numbers of qubits.
A particularly interesting approach to solving the scaling problem is networking of smaller machines via quantum coherent interconnects \cite{cirac1997quantum,duan2001long,kimble2008quantum}.
Communication between quantum processing units at microwave frequencies must however be carried out through cryogenically cooled waveguides \cite{magnard2020microwave} to avoid decoherence.
%have led to the pursuit of a plethora of strategies for scaling to larger numbers of qubits to create 
%today's laboratory-scale machines into universal computers with practical use that can outcompete classical counterparts. 
 An attractive alternative is to map the quantum state from the microwave domain to the optical domain, where information can be transmitted quantum coherently through low-loss optical fibers \cite{landry2007quantum}. 
Numerous methods have been  proposed  for microwave-to-optical conversion \cite{lauk2020perspectives,lambert2020coherent,zhong2020proposal,rueda2019electro,soltani2017efficient,rueda2016efficient,fan2018superconducting,witmer2020silicon,javerzac2016chip,hisatomi2016bidirectional,vogt2019efficient,han2018coherent,bartholomew2020chip,everts2019microwave,williamson2014magneto,tsang2010cavity,tsang2011cavity}; the most efficient to date make use of an intermediary mechanical state \cite{andrews2014bidirectional,bagci2014optical,higginbotham2018harnessing,jiang2020efficient,balram2016coherent,vainsencher2016bi,bochmann2013nanomechanical,shao2019microwave}. 

Great strides have been made in the field of cavity optomechanics \cite{aspelmeyer2014cavity} during the past decade in demonstrating the building blocks needed for quantum-coherent optical interconnects, including the demonstration of quantum control of single-phonon states in mechanical modes at gigahertz frequencies with single photons \cite{riedinger2016non}, the realization of optomechanically mediated quantum entanglement of separate mechanical systems \cite{riedinger2018remote}, and the conversion of a superconducting qubit excitation to an optical photon \cite{mirhosseini2020superconducting}.  
Progress has also been made in the manipulation of quantum states of a mechanical oscillator through interaction with a superconducting qubit
%circuit QED system 
\cite{chu2018creation,satzinger2018quantum,arrangoiz2019resolving}. 

We introduce here a novel platform for microwave-optical conversion that 
%allows us to 
combines a microwave circuit fabricated on an intrinsic silicon substrate with an optomechanical element made of 
%fabricated from 
single-crystal gallium phosphide (GaP), using direct wafer bonding. 
%The piezoelectric properties of GaP allow us to actuate the mechanical modes of the optomechanical devices with integrated electrodes and coherently transduce a microwave signal to an optical signal. 
For the optical cavity, we employ a quasi-one-dimensional optomechanical crystal cavity \cite{chan2011laser} with extended mechanical modes that permit electrical actuation remote from the optical mode to reduce losses. The extended modes nevertheless maintain substantial optomechanical coupling.
%spatially separated coupling to microwave and optical modes. This serves to both minimize losses and at the same time maintain substantial optomechanical and electromechanical coupling strengths.
%GaP possesses an attractive combination of physical properties.
% the wide electronic bandgap greatly reduces two-photon absorption at telecommunication wavelengths and 
%several unique properties, such as a large electronic band gap, high refractive index, and a significant piezoelectric response.
Due to the relatively high index of refraction of GaP (3.05 at $\lambda_{vac}=1550$ nm) 
and the consequent strong light confinement, GaP 
%optomechanical crystal 
cavities have vacuum optomechanical coupling rates similar to those of  devices made of silicon \cite{schneider2019optomechanics,stockill2019gallium} and significantly larger than  those of lithium niobate \cite{jiang2020efficient} or aluminum nitride \cite{bochmann2013nanomechanical} devices.
%Recent experiments with GaP photonic crystal cavities fabricated from gallium phosphide have shown similar properties with regard to optomechanical coupling rate and delayed heating at single phonon energies to those found in devices fabricated from silicon \cite{stockill2019gallium}.
%GaP devices compare favorably 
In terms of heating in pulsed experiments, GaP devices compare favorably  to structures made of 
gallium arsenide and other piezoelectric materials \cite{stockill2019gallium,forsch2020microwave}. %because the wider electronic bandgap of GaP
%due to its greater optical bandgap
%mitigates two-photon absorption. 
%, while providing a larger optomechanical coupling compared to devices made of lithium niobate \cite{jiang2020efficient} or aluminum nitride \cite{bochmann2013nanomechanical}. 
Making use of the piezoelectric properties of GaP, we demonstrate at room temperature actuation of the mechanical modes of the optomechanical device with integrated electrodes and coherently transduce microwave signals to optical frequencies. 

\section*{Device Design}
\begin{figure*}[t]
	\includegraphics[width = 2.1\columnwidth]{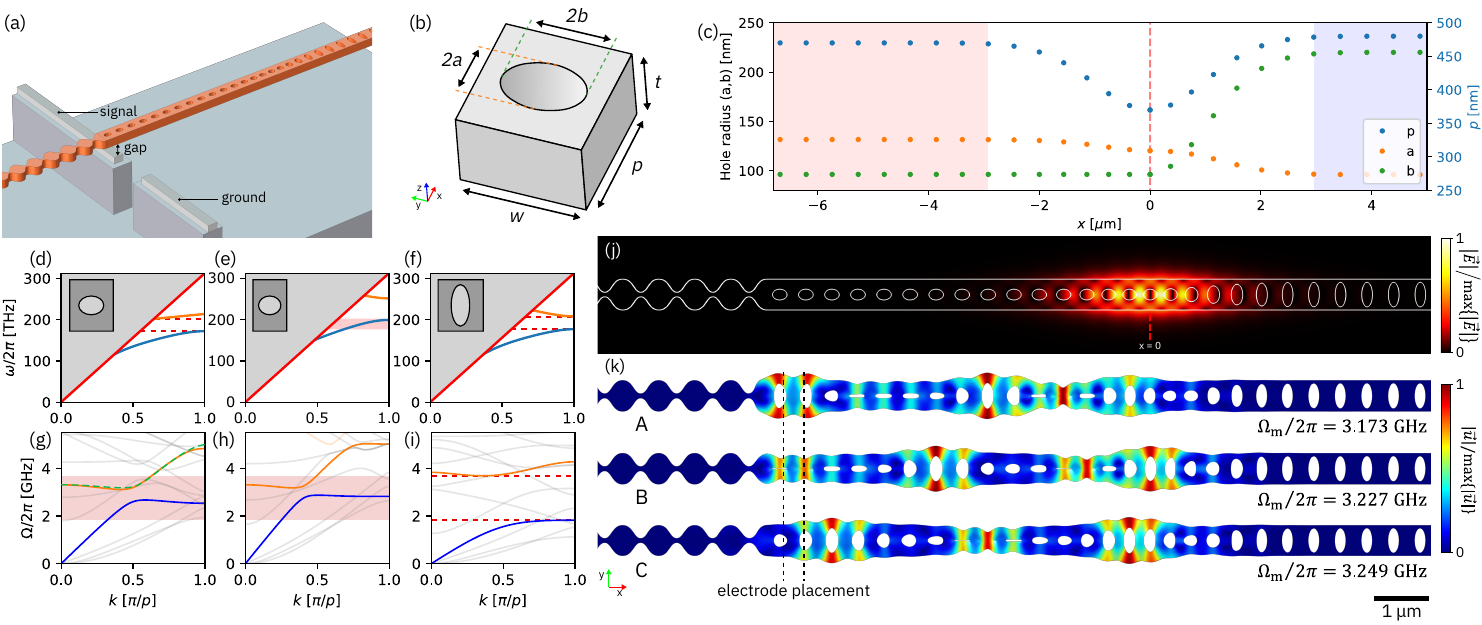}
	\caption{\color{black}\textbf{GaP optomechanical crystal cavity design.} (a) Illustration of the cavity suspended over an electrode pair. (b) Schematic of unit cell ($t = \SI{300}{nm}$, $w = \SI{550}{nm}$).  
	(c) Variation of unit cell parameters as indicated in (b) along the cavity x axis. The shading indicates the phononic-waveguide (red), cavity (white), and input-mirror (blue) regions.
	(d,e,f) Photonic band structure	of  the unit cell for the phononic waveguide, cavity center, and input mirror, respectively. The light line is indicated in red. The red shaded region in (e) is the intersection of the photonic bandgaps in (d) and (f).
	(g,h,i) Phononic band structure of the unit cell ordered as in (d,e,f).
	The mechanical breathing modes are indicated in blue and orange.  The red shaded region in (g) and (h) corresponds to the phononic bandgap in (i). 
	For comparison, the upper breathing mode band for the unit cell at the cavity center is overlayed in (g) as a dashed green line scaled for the difference in period $p$.
	(j) FEM simulation of the localized optical mode. The color scale indicates the magnitude of the electric field $|\vec{E}|$. 
	(k) FEM simulation of 
	%four 
	three extended mechanical breathing modes. The color scale indicates the magnitude of the mechanical displacement $|\vec{u}|$.}
	\label{sim_fig_2}
	\vspace{-3mm}
\end{figure*}

Our transducer consists of an asymmetric quasi-one-dimensional optomechanical crystal cavity made of GaP suspended over a niobium electrode at one end, with a second, coplanar niobium electrode displaced to one side, as illustrated in \fref{sim_fig_2}(a).
Laser light is coupled in and out of the device through a waveguide attached to it on the right side. 
The geometry  of the optomechanical crystal cavity is adapted from the design
%We adapt a photonic crystal design with elliptical holes etched into a rectangular beam as
introduced by Chan et al. \cite{chan2011laser,chan2012optimized} and consists of a series of elliptical holes etched in a rectangular beam.
%The geometry is as illustrated in \fref{sim_fig_2} (a). 
The hole radii and pitch are however modified to create distributed mechanical modes extending between the center of the cavity and the electrode position  
%Recently, a similar design was also implemented in  
%in a fashion similar to the device described by Mirhosseini et al. 
\cite{mirhosseini2020superconducting}.   
Finite-element-method (FEM) simulations are employed to model both the photonic and phononic band structure, determine the intrinsic, radiation-limited optical quality factor, $Q_0$, and estimate the vacuum optomechanical coupling rate, $g_0$. 

The design is composed of three sections (\fref{sim_fig_2}(c)). 
At the right end of the GaP beam (which is \SI{300}{nm} thick and \SI{550}{nm} wide), the unit cell is chosen to form a partially transparent photonic mirror with 
%The input (left in panel (i) of \fref{sim_fig_2}) supports 
an optical bandgap for TE-polarized light between \SI{177}{THz} and \SI{208}{THz}, where the optical coupling rate to the cavity can be controlled by varying the number of holes.  This section also has a bandgap for mechanical breathing modes between \SI{1.82}{GHz} and \SI{3.69}{GHz} and thus serves as a phononic mirror as well.
%at $x=0$ by a defect in the photonic crystal following a Gaussian function with a width $\sigma = \SI{2}{\micro m}$, where the structure supports a TE-polarized optical mode within the bandgap of the photonic crystal as well as a mechanical breathing mode. 
On the left side of the device, the unit cell geometry is again chosen to create a photonic bandgap, this time between \SI{172}{THz} and \SI{202}{THz}, but now has a phonon dispersion that supports mechanical breathing modes within the phononic bandgap of the input mirror on the right side of the device.
% the photonic crystal geometry is modified such that it supports a photonic bandgap between \SI{171.9}{THz} and \SI{202.0}{THz}, while crucially also supporting 
Between these two regions, in the central portion of the beam, the unit-cell dimensions are varied to produce a confined optical mode, with the variations on each side following Gaussian functions of  width $\sigma = \SI{1}{\micro m}$.  The result is a cavity supporting a localized optical mode (\fref{sim_fig_2}( j)) at $\omega_\textrm{o}/2\pi = \SI{193.2}{THz}$ with 
%an intrinsic, radiation-loss-limited quality factor 
$Q_0=1.8\times 10^6$. 
Crucially, the dispersion of the mechanical breathing modes is nearly identical 
%for the unit cells 
at the center of the cavity 
%region have nearly the same frequencies and wavevectors as those for the unit cell 
and on the left side of the device (see the dashed green line in \fref{sim_fig_2}(g)), resulting in an impedance-matched phononic waveguide.
%the ones supported by 
%For illustration the upper breathing mode band of the cavity cell is shown as dashed green line in \fref{sim_fig_2} (h) (rescaled to account for different period $p$). 
%A localized optical mode is found at $\omega_\textrm{o}/2\pi = \SI{193.2}{THz}$ with an intrinsic, radiation-loss limited quality factor $Q_{i,\textrm{rad}}=1.8\times 10^6$. 
%A mechanical breathing mode with $\Omega_\textrm{m}/2\pi = \SI{3.249}{GHz}$ co-localized with the optical mode is found with an estimated vacuum optomechanical coupling rate $g_0/2\pi = \SI{515}{kHz}$. We note here that this coupling rate is reduced compared to a design with stronger localization of the mechanical mode at the optical cavity. The localized optical mode and the mechanical mode are shown in \fref{sim_fig_2} (m) and  (n,o), respectively.
%The photonic crystal 
The device is terminated on the left with a phononic reflector exhibiting a complete bandgap between \SI{2.84}{GHz} and \SI{3.51}{GHz}, implemented as a crenulation of the nanobeam.
%, as shown in \fref{sim_fig_2}(o).
This reflector, in combination with the phononic mirror on the right side of the device, forms a Fabry-Pérot-like cavity for mechanical breathing modes.
% in the photonic crystal. 
Examples of the mechanical breathing modes are shown in \fref{sim_fig_2}(k), the frequencies and estimated vacuum optomechanical coupling rates of which are given in \autoref{mechanical_modes}.
%$\Omega_\textrm{m}/2\pi = \SI{3.249}{GHz}$ 
%$g_0/2\pi$ = \SI{515}{kHz}$ 
Because the mechanical breathing modes extend along the beam, they can be directly actuated by electrodes that are spatially separated from the localized optical mode. 
%Optical coupling to the system is achieved via a waveguide connected to the input mirror of the photonic crystal. 
%The coupling rate is increased by reducing the number of holes to the left of the cavity. 
\\
\setlength{\tabcolsep}{8pt}
\begin{table*}[p]
	\centering
	\vspace{-6mm}
	\caption{Simulated and measured parameters for the dominant mechanical breathing modes.}\vspace{2mm}
	\begin{tabular}{|c|c|c|c|c|c|c|c|c|c|}
		\hline
		& \multicolumn{2}{c}{Simulated} &&  \multicolumn{5}{c}{Measured}&\\
		\hline
		\multirow{2}{*}{Mode}& $\Omega_\textrm{m}/2\pi$& $g_0/2\pi$& $\Gamma_\textrm{ex}/2\pi$ & $\Omega_\textrm{m}/2\pi$ & \multirow{2}{*}{$Q_\textrm{m}$} & $g_0/2\pi$ [kHz]& $g_0/2\pi$ [kHz]&  $\Gamma_\textrm{ex}/2\pi$& $\phi_\textrm{m}/2\pi$  \\
		& [GHz]& [kHz]& [mHz]&[GHz] & & (blue detuned)& (red detuned) & [mHz] & [rad]\\ 
		\hline\hline
		A& 3.173& 173& 6.20& 3.280& 1288 &193& 189& 2.32& 0\\
		\hline
		B& 3.227& 381& 8.02& 3.314& 1180 &285& 281& 3.80& $\pi$\\
		\hline
		C& 3.249& 506& 25.2& 3.328& 1301 &294& 291& 1.75& 0\\
		%\hline
		%D& 3.315& 619& 1.11& 3.332& 215& 212& 0.33& 0\\
		\hline
	\end{tabular}
	\label{mechanical_modes}
\end{table*}
\begin{figure*}[t]
\includegraphics[width=2.05\columnwidth]{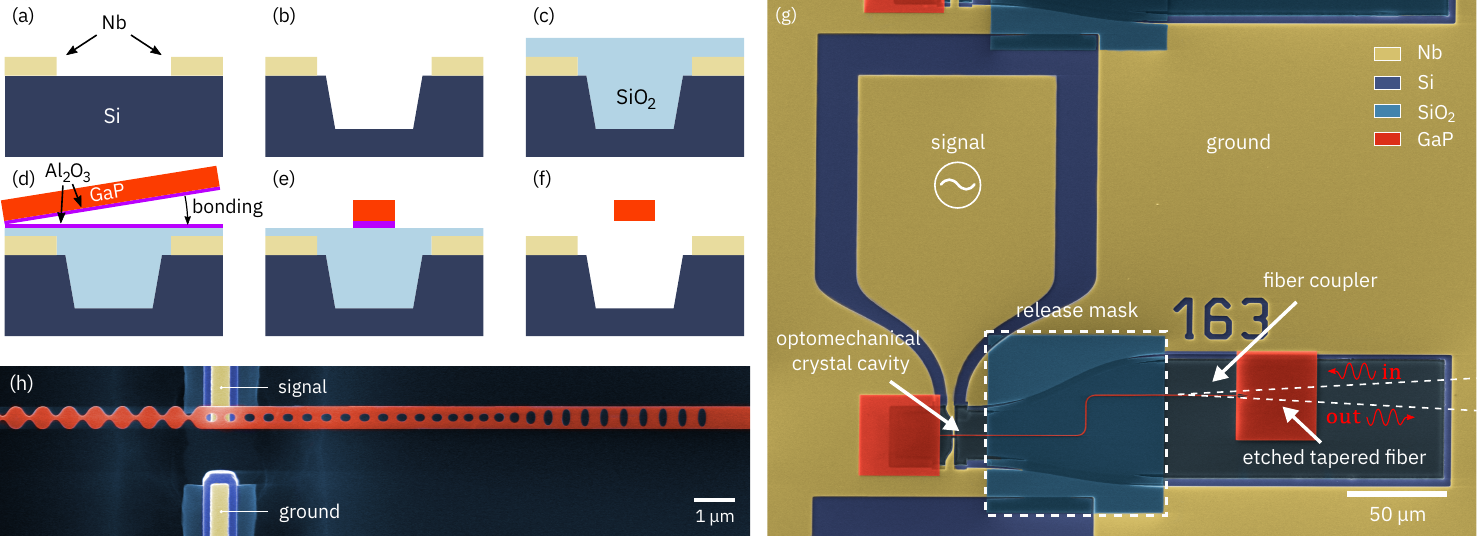}
\caption{\color{black}\textbf{Device fabrication process. 
%Fabrication of the GaP photonic crystal cavity on a Si wafer with Nb electrodes.
} (a) Deposition and patterning of Nb electrodes. (b) Local recessing of the  Si substrate. (c) Deposition and planarization of the sacrificial SiO$_2$ layer. (d) Direct wafer bonding of the GaP device layer. (e) Patterning of the integrated GaP photonic circuit. (f) Release of the free-standing portion of the device. (g) Scanning electron microscope (SEM) image of the finished device with annotation of microwave and optical ports. (h) Close-up SEM image of the optomechanical crystal cavity.}
\label{fab_fig}
\vspace{-3mm}
\end{figure*}
\section*{Fabrication}
Device fabrication makes use of the processes described in our previous work on integrated GaP photonics
%Fabrication of the device illustrated in \fref{sim_fig_2} follows the process we described in earlier work 
\cite{schneider2019optomechanics,wilson2020integrated}, 
%Crucially,
with the important distinction that GaP
%the piezoelectric crystal
is not bonded onto an oxidized silicon wafer but instead onto a \textit{pre-fabricated} microwave circuit.  The process flow is illustrated in \fref{fab_fig}.
First, a \SI{250}{nm}-thick niobium film is deposited on a 
%4-inch 
intrinsic silicon wafer by magnetron sputtering and patterned by chlorine-based dry etching into electrode structures.
%Instead, microwave circuits are first fabricated on an intrinsic silicon substrate. 
The vicinity of the photonic crystal cavity is then recessed $\SI{2}{\micro m}$ by dry etching of the silicon with a HBr/O$_2$ mixture to minimize optical losses to the substrate. The wafer is then covered with a sacrificial SiO$_2$ layer and the surface planarized for bonding of GaP. The thickness of the SiO$_2$ layer determines the eventual gap ($\sim\SI{300}{nm}$) between the GaP beam and the underlying niobium electrode.  After dicing into chips, the GaP device layer is patterned by electron-beam lithography to form the photonic crystal cavity and the attached waveguide, which has a tapered end for adiabatic coupling to a fiber with a tapered tip formed by etching. The freestanding portion of the device is defined photolithographically and released by removal of the sacrificial SiO$_2$ layer with buffered HF.  Finally, the entire chip is coated with \SI{8} {nm} of Al$_2$O$_3$  to protect the surface and prevent photooxidation during measurement.
%The process is outlined in \fref{fab_fig} with a SEM micrograph of a finished device shown in panel (g), more 
Further details, including  a description of the adiabatic fiber coupling to the device, are given in the Supplementary Information. 

\section*{Optomechanical Characterization}
%The device is characterized at room temperature with the apparatus shown in \fref{exp_fig1}(a).
The optical cavity mode is first interrogated in reflection with a tunable diode laser at low power ($
%P_\textrm{opt}
\sim \SI{130}{nW}$)
%A tunable laser is first used for characterization of the optical cavity signal in reflection 
to determine the resonance frequency $\omega_\textrm{o}$ and the loaded quality factor $Q$.
\begin{figure}[t]
	\includegraphics[width=0.9\columnwidth]{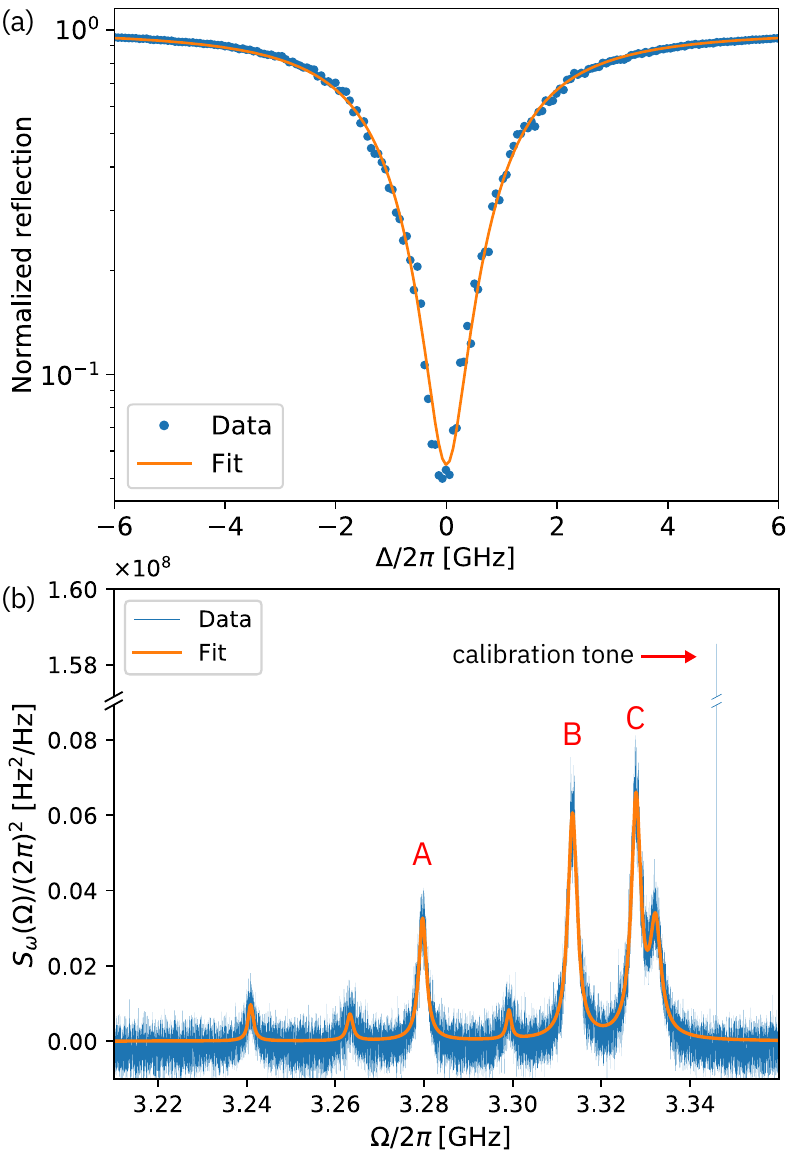}
	\caption{\color{black}\textbf{Optical and thermo-mechanical spectroscopy}: 
		%(a) Schematic of characterization apparatus.  ECDL, tunable external cavity diode laser; VOA, variable optical attenuator; FPC, fiber polarization controller; PM, phase modulator; EDFA, erbium-doped fiber amplifier; BP, bandpass filter; ESA, electrical spectrum analyzer; OSA, optical spectrum analyzer; VNA, vector network analyzer; DUT, device under test. (b) Photograph of chip during measurement with aligned tapered fiber and microwave probe. (c) Optical micrograph through alignment microscope showing the device. (d) SEM image of the photonic crystal cavity. 
		(a) Fit (orange) of equation (\ref{eq. reflection})  to the optical reflection spectrum (blue) of the cavity resonance. 
		(b) Thermo-mechanical spectrum (blue) plotted as symmetrized frequency noise spectral density $S_\omega(\Omega)$ with a fit to a sum of Lorentzian functions (orange). The data is calibrated with an added tone created by a phase modulator. The assignment to the modes shown in \fref{sim_fig_2}(k) is indicated by A, B, and C.} 
	%The blue curve shows filtered (Savitzky-Golay) data as a guide to the eye.}
	\label{exp_fig1}
	%\vspace{-3mm}
\end{figure}
For a Lorentzian resonance, reflection from the cavity
%The cavity reflection spectrum is shown in \fref{exp_fig1} (e) in blue 
as a function of detuning $\Delta = \omega_\textrm{L}-\omega_\textrm{o}$ with respect to the laser frequency $\omega_\textrm{L}$ is described by 
%With a fit to the Lorentzian cavity reflection 
\begin{equation}
	\label{eq. reflection}
	|\mathcal{R}(\Delta)|^2 = \left|\frac{(\kappa_0-\kappa_\textrm{ex})/2-i\Delta}{(\kappa_0+\kappa_\textrm{ex})/2-i\Delta}\right|^2,
\end{equation}
where $\kappa_0$ is the intrinsic cavity decay rate  and $\kappa_\textrm{ex}$ is the external coupling rate \cite{aspelmeyer2014cavity}.  A fit to \eqref{eq. reflection} of the resonance observed at $\omega_\textrm{o}/2\pi = \SI{196.5}{THz}$ (\fref{exp_fig1}(a)) gives a loaded optical quality factor $Q = \omega_\textrm{o}/\kappa = 6.73 \times 10^4$. 
The coupling factor is $\eta_\textrm{c,opt} = \frac{\kappa_\textrm{ex}}{\kappa_\textrm{ex}+\kappa_0} = 0.383$, yielding an intrinsic quality factor $Q_0= 1.09\times 10^5$.  Similar values have been achieved with GaP optomechanical crystal cavities reported in other work \cite{schneider2019optomechanics,stockill2019gallium}. 
In addition to surface roughness and fabrication imperfections, absorption in the bulk of the GaP or at the surface may be limiting the intrinsic quality factor.
We also find an optical mode frequency that is \SI{1.7}{\percent} higher than expected from simulations, which we attribute to small deviations from the design geometry. 

\begin{figure*}[t!]
	\includegraphics[width=1.8\columnwidth]{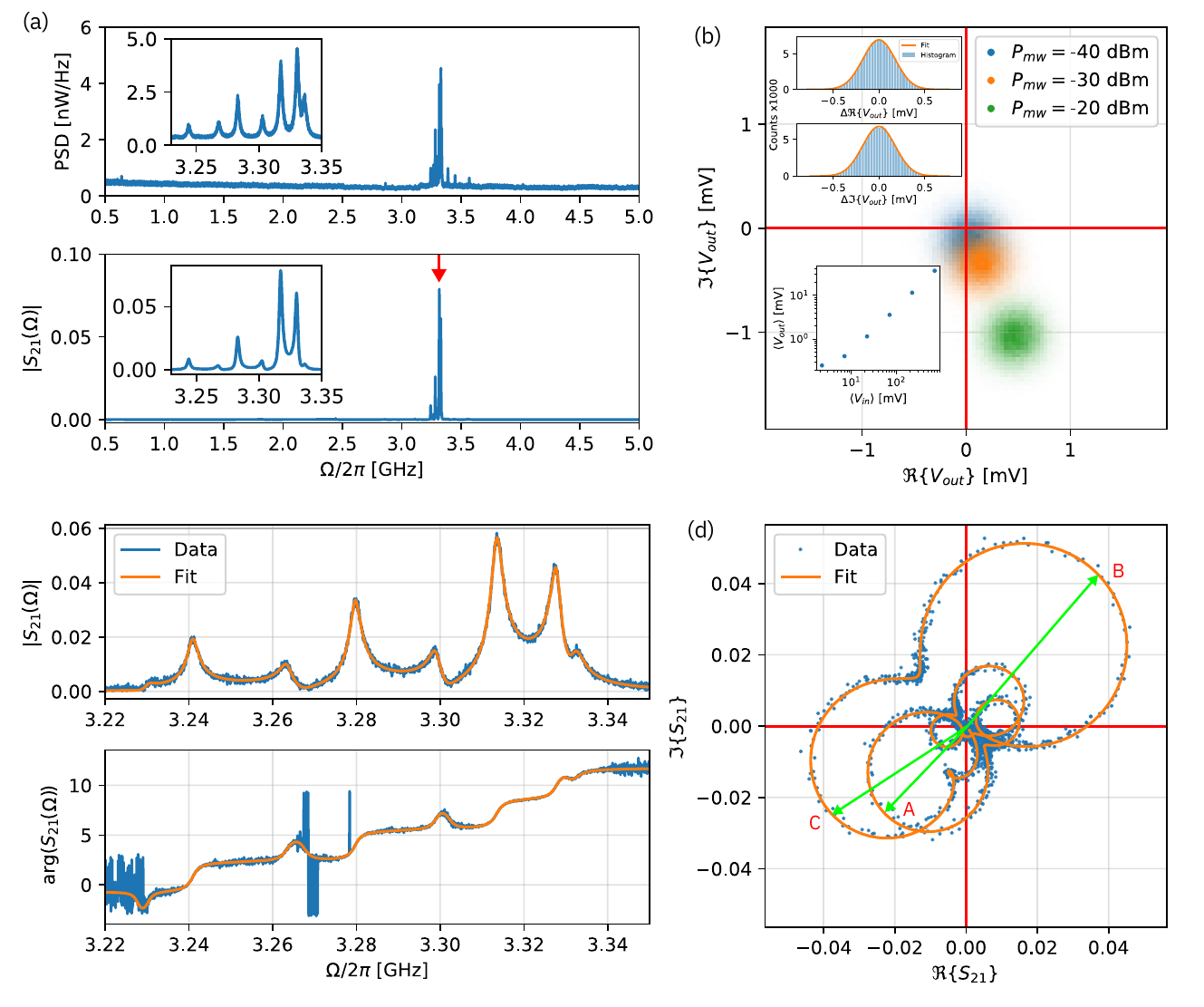}
	\caption{\color{black}\textbf{Microwave-to-optical transduction.} (a) Thermo-mechanical noise power spectral density (PSD) (top) and microwave-to-optical transmission amplitude $|S_{21}|$ (bottom). (b) Quadrature components of the demodulated photoreceiver output voltage distribution $V_\textrm{out}$ when the device is driven at \SI{3.317}{GHz} (indicated by red arrow in the bottom panel of (a)) at three microwave powers. For each power level, $10^5$ points were collected at a bandwidth of \SI{10}{kHz}. The insets show histograms of the real and imaginary part of the noise distribution at $\SI{-20}{dBm}$ along with Gaussian fits (upper left) and the average output voltage $\langle V_\textrm{out}\rangle$ vs. the average microwave input voltage $\langle V_\textrm{in}\rangle$ (lower left). (c)  Amplitude and phase of the transduced signal versus drive frequency, where the contribution due to cable delay $\tau$ has been removed. (d) In-phase versus quadrature component of the transduced signal. Actuation at the measured eigenfrequencies of the dominant modes is indicated by green phasors, with the assignment to the modes shown in \fref{sim_fig_2}(k) indicated by A, B, and C.}
	\label{exp_fig2}
	\vspace{-3mm}
\end{figure*}

Thermally driven mechanical motion of the photonic crystal cavity modulates the cavity resonance frequency, which, for a detuned probe laser, produces amplitude fluctuations in the reflected light. We measure the resulting thermo-mechanical spectrum 
%from \SI{1}{GHz} to \SI{5}{GHz} 
by direct detection with a fast photoreceiver.
%for both blue and red laser detuning. 
%to identify optomechanically coupled mechanical modes. 
%We found the 
At frequencies above \SI{500}{MHz}, the only modes with appreciable optomechanical coupling are observed in a narrow band between \SI{3.21}{GHz} and \SI{3.35}{GHz} (\fref{exp_fig1} (b) and \fref{exp_fig2}(a)), in good agreement with the simulated mechanical breathing modes. 
%The broadband thermo-optomechanical spectrum is shown in the top panel of \fref{exp_fig2} (a). 
As the total optical decay rate is measured to be $\kappa=\kappa_0+\kappa_\textrm{ex}= 2\pi\times\SI{3.01}{GHz}$, the system is narrowly in the resolved-sideband regime.

The vacuum optomechanical coupling rate $g_0$ of individual modes was determined via noise calibration \cite{gorodetksy2010determination}. 
To that end, the thermo-mechanical spectrum is measured at low cooperativity ($\mathcal{C}=\bar{n}_\textrm{cav}\frac{4g_0^2}{\Gamma_{\textrm{m}} \kappa} \approx 4\times 10^{-3}$, where $\Gamma_{\textrm{m}}$ is the mechanical damping rate and $\bar{n}_\textrm{cav}$ is the average number of photons in the cavity), so that effects due to dynamical backaction can be neglected.  A reference signal generated by a phase modulator is added close to the mechanical modes to calibrate the spectrum. 
The modulation depth $\phi_0 = {\pi V_0}/{V_\pi}$ (where $V_0$ is the amplitude of the radio-frequency voltage applied to the modulator and $V_\pi$ is the half-wave voltage) at frequency $\Omega_\textrm{mod}$ is determined from the first sideband ratio (see Supplementary Information). 
The vacuum optomechanical coupling rate is then calculated as \cite{gorodetksy2010determination}
\begin{equation}
	g_0 \approx \frac{\phi_0}{2}\Omega_\textrm{m}\sqrt{\frac{1}{\langle n_\textrm{th}\rangle}\frac{S_\textit{V}(\Omega_\textrm{m})\Gamma_\textrm{m}/4}{S_\textit{V}(\Omega_\textrm{mod})f_\textrm{ENBW}}},
	\label{g0_gor}
\end{equation}
where 
%$\phi_0 = \frac{V_0\pi}{V_\pi}$ is the modulation depth ($V_0$ is the amplitude of the RF voltage applied to the modulator), 
$\langle n_\textrm{th}\rangle\approx \frac{k_B T}{\hbar \Omega_\textrm{m}}$ is the average thermal occupation of the mode, $S_\textit{V}(\Omega)$ is the symmetrized voltage noise spectral density produced by the photoreceiver, $\Gamma_\textrm{m}$ is the mechanical damping rate, and $f_\textrm{ENBW}$ is the effective noise bandwidth of the spectrum analyzer's filter function. 
We fit the mechanical noise spectrum with a sum of Lorentzians, as each mode produces a voltage noise spectral density that is uncorrelated with the other modes. 

The device exhibits appreciable optomechanical coupling ($g_0/2\pi>\SI{100}{kHz}$) for 
%four 
three modes in the frequency range ascribed to the breathing modes, which we tentatively assign to the modes depicted in \fref{sim_fig_2}(k). 
%with vacuum optomechanical coupling rates $g_0/2\pi>\SI{100}{kHz}$. 
The observed mechanical  resonance frequencies and coupling rates are listed in \autoref{mechanical_modes}. We attribute the higher frequency of the experimentally observed modes to the Al$_2$O$_3$ protective coating, which increases the stiffness of the nanobeam (see Supplementary Information).  The noise calibration measurement was carried out at blue and red laser detuning;
% to verify that dynamical backaction is negligible. The 
the coupling rates for blue detuning are slightly higher than for red detuning, consistent with a small contribution from residual dynamical backaction. 
%We also note here that due to the overlap between the modes at \SI{3.328}{GHz} and \SI{3.332}{GHz} the measured current noise spectral density does of course not correspond to $S_\textit{II}(\Omega_\textrm{m})$ in \eqref{g0_gor} where we use the amplitude of the individual Lorentzian derived from the fit, as for all measured modes. 
%\\
%\begin{table}[h]
%	\centering
%	\vspace{-6mm}
%	\caption{Measured resonance frequencies $\Omega_\textrm{m}$ and vacuum optomechanical coupling rates $g_0$ of the dominant mechanical breathing modes.}\vspace{2mm}
%	\begin{tabular}{|c|c|c|}
%		\hline
%		$\Omega_\textrm{m}/2\pi$ [GHz]& $g_0/2\pi$ [kHz] (blue)& $g_0/2\pi$ [kHz] (red) \\
%		\hline\hline
%		3.280&  193& 189\\
%		\hline
%		3.314&  285& 281\\
%		\hline
%		3.328&  294& 291\\
%		\hline
%		3.332&  215& 212\\
%		\hline
%	\end{tabular}
%\label{caltone_table}
%\end{table}

\section*{Piezoelectric Coupling}
\begin{figure}[t!]
	\includegraphics[width=\columnwidth]{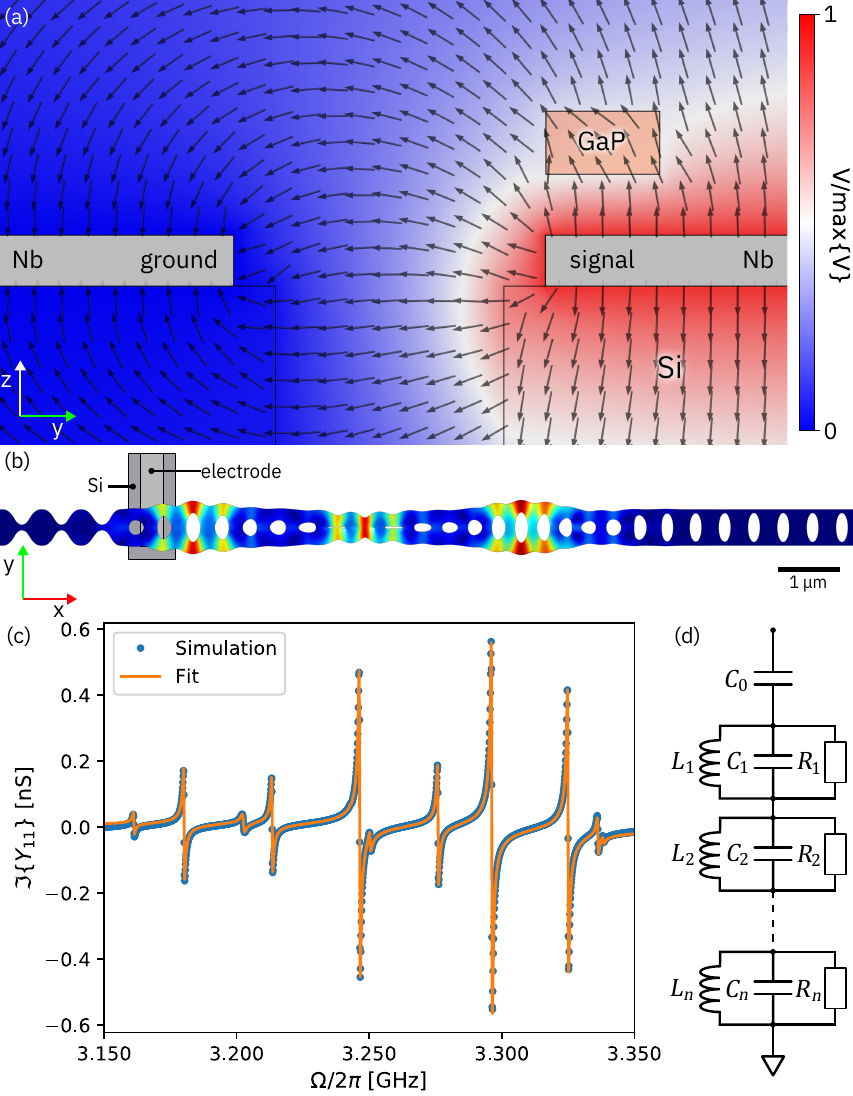}
	\caption{\color{black}\textbf{FEM simulation of piezoelectric coupling.} (a) Cross-section of the electric potential between the coplanar microwave signal (red) and ground (blue) electrodes. The black arrows indicate the electric field direction. The orange box is the GaP beam. 
	%(b) Illustration showing the location of the photonic crystal cavity with respect to the electrode pair. 
	(b) FEM simulation of mechanical breathing mode C showing the position of the signal electrode. (c) Imaginary part of the simulated piezoelectric admittance $Y_{11}$ (blue) and the fit to the LC network (orange). (d) Circuit representation of the series network of parallel LC resonators used for the admittance fit. 
	%(f) Circuit representation of Transmon qubit coupled to parallel LC resonator. (g) Illustration of mechanical breathing mode coupled in optimized electrode geometry. (h) Imaginary part of piezoelectric admittance in optimized geometry. (h) Calculated coupling to Transmon qubit (red) as function of $E_J/E_C$ and corresponding qubit shunt capacitance $C_\Sigma$.
	}
	\label{sim_fig}
	%\vspace{-3mm}
\end{figure}

We now turn the focus of the discussion to piezoelectrically mediated actuation of the optomechanical cavity 
%with an externally applied electric field. 
and analyze the frequency dependence of the coupling with a multiphysics FEM simulation in which the signal voltage is applied to the electrode directly under the GaP beam and the other electrode is ground (\fref{sim_fig}).
The zinc-blende crystal structure of GaP dictates that mechanical breathing modes are most effectively actuated with the axis of the optomechanical crystal cavity aligned with the $[011]$ direction of the GaP crystal lattice and the electric field oriented along the $[100]$ direction (z-axis in \fref{sim_fig}). 
As can be seen from \fref{sim_fig}(a), the electric field for our coplanar electrode geometry has a component along the z-axis but is in general oriented at an angle.
The coplanar electrode arrangement simplifies fabrication but is clearly not ideal.  An optimized geometry is discussed below.

The admittance $Y_{11}$ of the device as a function of frequency is calculated and fit to that of an equivalent network consisting of a series of parallel LC resonators \cite{arrangoiz2016engineering} (\fref{sim_fig} (e)). 
A small imaginary part is introduced to the elasticity tensor of GaP to broaden the spectral response of the individual modes and permit a coarser sampling
%allow for a lower sampling rate 
of the spectrum. 
The resulting added loss is modeled as parallel resistors in the LC elements. 
The imaginary part of the admittance 
%for the coplanar geometry 
is shown in \fref{sim_fig} (d) along with the fit to the LC network, where the 
electrostatic contribution of the coupling capacitor $C_0$ (which is determined by the geometry and dominates the admittance) has been subtracted for clarity. 
We find that the equivalent network provides an accurate representation of the electromechanical coupling. 

With an equivalent circuit model in place, we can estimate the coupling of each mechanical mode to the microwave probe used experimentally by employing a semi-infinite transmission-line model.  Specifically, the coupling capacitor $C_0$ is assumed to be connected to a load impedance $Z_0 = \SI{50}{\ohm}$, yielding for the $n$-th mode an electromechanical coupling rate of
\begin{equation}
\Gamma_{\textrm{ex},n} \approx \frac{Z_0 C_0^2 \Omega_{\textrm{m},n}^2}{C_n+C_0},
\end{equation}
with $\Omega_{\textrm{m},n}^2 = 1/L_n C_n$ (see Supplementary Information for details). 
The values simulated for $\Gamma_{\textrm{ex},n}$ with
%We find a maximum electromechanical coupling of $\Gamma_\textrm{ex}/2\pi = \SI{25.3}{mHz}$ for the mode shown in \fref{sim_fig} (c) and 
a coupling capacitance of $C_0 = \SI{0.42}{fF}$ are listed in \autoref{mechanical_modes}.
%We note here that we found a deviation between the eigenfrequencies extracted from the multiphysics simulation compared to a mechanical eigenmode solution of $\sim\SI{50}{MHz}$ which may be attributed to the different simulation setup. 

\section*{Microwave-to-Optical Transduction}
We demonstrate microwave-to-optical transduction 
%with the device shown in \fref{fab_fig}(g) 
by driving 
%the port connected to 
the electrode below the device with a vector network analyzer (VNA)
and measuring the optical signal from a detuned pump laser reflected from the cavity. 
%on the detuned pump laser in reflection from the cavity. 
The optomechanically induced modulation is observed by direct detection with a fast photoreceiver.
The results are shown in \fref{exp_fig2}. 
%The transduction spectrum is measured as the transmission between the microwave port of the device (port 1) and the photoreceiver that detects the optical signal (port 2). 
%We deploy direct detection to detect the optomechanically induced phase modulation. 
We evaluate the microwave-to-optical transduction by considering the system as a two-port network in which port 1 is the microwave probe and port 2 the photoreceiver. The scattering parameter $S_{21}(\Omega)$ as a function of the VNA frequency $\Omega$ is then given by 
\begin{equation}
	\begin{aligned}
		S_{21}(\Omega) = \frac{\Gamma_\textrm{D}}{\sqrt{2Z_0 \hbar \Omega}}  &\Theta(\Omega) e^{-i(\Omega\tau+\theta)} \\&\cdot\sum_n g_{0,n}\sqrt{\Gamma_{\textrm{ex},n}}\chi_{\textrm{m},n}(\Omega)e^{i\phi_{\textrm{m},n}},
	\end{aligned}
	\label{S21fct}
\end{equation}
as derived in the Supplementary Information. Here, $\Gamma_\textrm{D}$ denotes a detection gain factor that is determined by the optical pump power, the photoreceiver's quantum efficiency, as well as all electrical and optical loss and gain in the system. 
The subscript $n$ denotes the parameters for the n-th mechanical mode.
%$g_{0,n}$, $\Gamma_{\textrm{ex},n}$, and $\chi_{\textrm{m},n}(\Omega)$ denote the vacuum optomechanical coupling rate, the electromechanical transmission line coupling, and the mechanical susceptibility of the $n$-th mode, respectively, where 
The mechanical susceptibility is
\begin{equation}
	\chi_{\textrm{m},n}(\Omega) = \frac{1}{\frac{\Gamma_{\textrm{m},n}}{2}-i(\Omega-\Omega_{\textrm{m},n})}
\end{equation} 
with the mechanical damping rate $\Gamma_{\textrm{m},n}$ and eigenfrequency $\Omega_{\textrm{m},n}$. 
%and eigenfrequency $\omega_{\textrm{m},n}$. 
%We consider the 
The signal delay through the cables and optical fiber is described by a frequency dependent phase offset $\Omega\tau + \theta$, where $\tau$ is the propagation time through the signal path and $\theta$ is a fixed arbitrary phase offset. Additionally, there is a mode-dependent phase offset $\phi_{\textrm{m},n} \in \{0,\pi\}$, corresponding to the relative phase difference between the mechanical displacement at the electromechanical and the optomechanical coupling positions.
%associated with the effective wavelength of the mechanical mode (discussed further below).
%depending on the mechanical mode order that transduces the microwave signal.
The optomechanically induced phase modulation is transduced into amplitude modulation
%, that is 
detected by the photoreceiver 
according to the function 
\begin{equation}
\begin{aligned}
\Theta(\Omega) = i\kappa_\textrm{ex}\Big(&\left(1-\kappa_\textrm{ex}\chi_\textrm{o}^*(0)\right)\chi_\textrm{o}(0)\chi_\textrm{o}(\Omega)-
\\&\left(1-\kappa_\textrm{ex}\chi_\textrm{o}(0)\right)\chi_\textrm{o}^*(0)\chi_\textrm{o}^*(-\Omega)\Big),
\end{aligned}
\end{equation}
where 
%$\kappa_\textrm{ex}$ is the external coupling rate of the optical cavity and 
\begin{equation}
	\chi_\textrm{o}(\Omega) = \frac{1}{\frac{\kappa}{2}-i(\Delta+\Omega)}
\end{equation}
is the optical susceptibility.
% with the laser detuning  $\Delta = \omega_\textrm{L}-\omega_\textrm{cav}$ and the cavity decay rate $\kappa$.
The measured 
%broadband 
microwave-to-optical transmission amplitude $|S_{21}|$ 
%from \SI{500}{MHz} to \SI{5}{GHz} 
is shown in \fref{exp_fig2}(a),
%the lower panel of (a). For comparison, 
along with the thermo-mechanical spectrum for comparison.
%is shown in the upper panel of the figure. 
%The frequency range where we found mechanical modes with appreciable optomechanical coupling is shown in both panels as inset. We found a 
The clear correspondence between the spectra 
%optomechanically coupled modes and peaks in the microwave-to-optical transmission, which 
indicates that the transduction is indeed mechanically mediated. 

We verify the phase-coherence of the transduced signal by recording the distribution of the photoreceiver output voltage %at three different microwave input powers
%by carrying out a continuous-wave measurement 
at a fixed frequency of \SI{3.317}{GHz} (where the highest transmission magnitude was observed), as displayed in \fref{exp_fig2}(b). 
%The recorded voltage distributions measured by the VNA with \SI{10}{kHz} bandwidth for three different microwave input powers are shown in panel (b) for a microwave input power of \SI{-40}{dBm} (blue) and \SI{-20}{dBm} (orange). 
%We found in both cases a 
The symmetric Gaussian distribution, which is dominated by thermo-mechanical noise,
%with a variance of $\langle \Re\{V_\textrm{out}\}^2\rangle = \langle \Im\{V_\textrm{out}\}^2\rangle = \SI{0.033}{mV\squared}$ and $\langle \Re\{V_\textrm{out}\}^2\rangle = \langle \Im\{V_\textrm{out}\}^2\rangle = \SI{0.032}{mV\squared}$, respectively. 
%We found a 
and the linear dependence on microwave input power of the displacement in phase space at constant phase 
%correspondence between the displacement at low power and high power, while the phase was preserved 
%The distributions in (b) are displayed with a \SI{90}{\degree} phase offset for clarity. 
%We therefore concluded that the transduction is indeed linear and coherent. 
confirm that the transduction is linear and coherent.

%The optical measurement setup was then calibrated by phase-modulating the optical pump with a pre-determined modulation depth and recording the cavity response (shown in \fref{exp_fig2}). 
%Using the previously measured vacuum optomechanical coupling rates, the electromechanical coupling rate $\Gamma_\textrm{ex}$ can then be determined via the microwave-to-optical transmission. 
The amplitude and phase of the transmission spectrum in the relevant range from \SI{3.22}{GHz} to \SI{3.35}{GHz} are shown in \fref{exp_fig2}(c) together with a fit to \eqref{S21fct}. 
The phase-space representation of the same data (\fref{exp_fig2}(d)), exhibits circular curves characteristic of harmonic oscillators. 
The fit makes use of a separate calibration of the optical cavity response to determine $\Gamma_\textrm{D}$, performed by phase-modulating the optical pump with a pre-determined modulation depth (see Supplementary Information).  Using the previously measured vacuum optomechanical coupling rates, the electromechanical coupling rate $\Gamma_\textrm{ex}$ is then determined from the microwave-to-optical transmission. 
%We identified 
%The three dominant mechanical modes that mediate the transduction at \SI{3.280}{GHz} (g), \SI{3.313}{GHz} (h) and \SI{3.328}{GHz} (i), that correspond to three of the four modes for which we previously calibrated the optomechanical coupling. 
%The phasors of the transduced signal at each frequency are indicated in (e) by green arrows. 
The fit results for the three mechanical modes with the strongest transduction, which we assign to modes A, B, and C, are shown in \autoref{mechanical_modes}.
%We find only the three modes shown in \fref{exp_fig2}(g,h,i) in the simulated structure that match the observed spectrum at \SI{3.246}{GHz}, \SI{3.276}{GHz} and \SI{3.296}{GHz}, respectively.
%We attribute the lack of significant transduction by mode D to its small mechanical displacement amplitude in the vicinity of the electrodes.
%We attribute the higher frequency of the experimentally observed modes to the Al$_2$O$_3$ protective coating, which increases the stiffness of the nanobeam.  

%\begin{table}[h]
%	\centering
%	\vspace{-4mm}
%	\caption{Eigenfrequencies $\Omega_\textrm{m}$ and vacuum optomechanical coupling rates $g_0$ of dominant mechanical modes.}\vspace{2mm}
%	\begin{tabular}{|c|c|c|}
%		\hline
%		$\Omega_\textrm{m}/2\pi$ [GHz]& $\Gamma_\textrm{ex}/2\pi$ [mHz]& $\phi_\textrm{m}/2\pi$ [rad] \\
%		\hline\hline
%		3.280&  2.32& 0\\
%		\hline
%		3.313&  3.80& $\pi$\\
%		\hline
%		3.328&  1.75& 0\\
%		\hline
%	\end{tabular}
%	\label{trans_table}
%\end{table}

Concerning the observed mode-dependent phase offset $\phi_\textrm{m}$ in the signal 
%as indicated in \eqref{S21fct} 
that we fix at either $0$ or $\pi$, we understand this phenomenon to be a result of the shape of each breathing mode's displacement field. 
When a mode is driven coherently, 
%at the microwave port on resonance,
the mechanical displacement field  in the portion of the GaP beam
%at the antinode 
that overlaps with the electrode inherits its relative  phase from the microwave drive. 
The optical mode however couples to a different region that might be in phase or out of phase by $\pi$ with respect to the piezoelectrically coupled region (see \fref{sim_fig_2}(k)). 
Consequently, the optomechanically induced phase modulation of the optical field may also have an added phase offset of $0$ (modes A and C) or $\pi$ (mode B) with respect to the microwave drive. 
Note that, although the device possesses several other breathing modes, only the three modes A, B, and C provide
%We find 
a reasonable qualitative agreement between simulations and experiment with respect to mechanical frequency, optomechanical coupling rate, and mode-dependent phase offset, justifying the assignment.
% only for the three modes depicted in \fref{sim_fig_2}(k) and assigned these modes to the observed ones, accordingly.

\begin{figure}[t!]
	\includegraphics[width=\columnwidth]{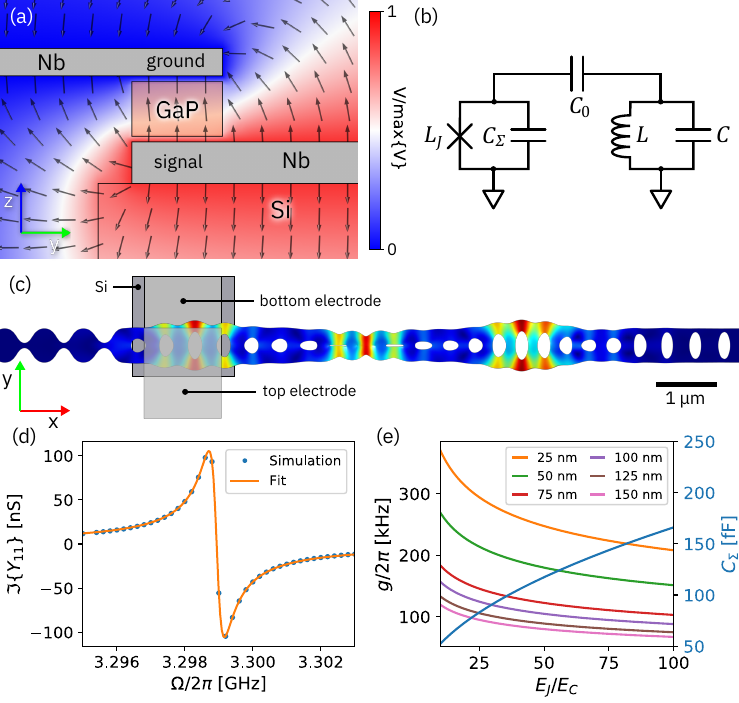}
	\caption{\color{black}\textbf{Theoretical analysis of coupling to a transmon qubit.} (a) Cross-section of a FEM simulation of the electric potential between the microwave signal (red) and ground (blue) electrodes for the optimized geometry. The black arrows indicate the electric field direction. The orange box is the GaP beam. (b) Circuit representation of a transmon qubit coupled to a parallel LC resonator. (c) FEM simulation of a mechanical breathing mode showing the electrode geometry. (d) Imaginary part of the simulated piezoelectric admittance $Y_{11}$ (blue) and the fit to the LC resonator (orange). (e) Calculated qubit shunt capacitance $C_\Sigma$ and coupling rates to a transmon qubit as function of $E_J/E_C$ for various gap sizes between the electrodes and the photonic crystal cavity.}
	\label{sim_fig3}
	\vspace{-3mm}
\end{figure}

%\section*{Transduction Efficiency}
%We may calculate the total transduction efficiency of the system as \cite{jiang2020efficient}
%\begin{equation}
%	\eta = \eta_{\textrm{c},\textrm{mw}}\eta_{\textrm{c},\textrm{opt}} \frac{4\mathcal{C}}{(1+\mathcal{C})^2},
%\end{equation}
%given the optomechanical cooperativity $\mathcal{C} = n_\textrm{cav}\frac{4 g_0^2}{\kappa\Gamma_\textrm{m}}$ and the external coupling factors to the system, $\eta_{\textrm{c},\textrm{mw}}=\frac{\Gamma_\textrm{ex}}{\Gamma_\textrm{m}}$ and $\eta_{\textrm{c},\textrm{opt}}=\frac{\kappa_\textrm{ex}}{\kappa}$. With $\mathcal{C}=4\times 10^{-3}$, at an optical pump power of $P_\textrm{opt} = \SI{130}{nW}$, we find the maximum transduction efficiency $\eta = 7.92 \times 10^{-12}$, limited by the microwave external coupling factor $\eta_{\textrm{c},\textrm{mw}} = 1.35\times 10^{-9}$.

\section*{Outlook:  Coupling to a Transmon Qubit}
%\subsection*{Coupling to a Transmon Qubit}
In addition to transmission line coupling, we also estimated the expected electromechanical coupling rate to a superconducting transmon qubit. 
We restrict the discussion here to an optimized configuration with one electrode below and the other above the photonic crystal cavity with equal gaps to the electrodes (\fref{sim_fig3}(a)).
%, as this is the likely geometry we will employ for this experiment in the future.  
The capacitive qubit-resonator coupling scheme is depicted in \fref{sim_fig3}(b), with the circuit parameters $C_0$, $L$ and $C$ extracted from an admittance fit similar to that described above.
%of the ideal geometry shown in panel (h) of the figure. 
In this case, we simulate only the admittance of mechanical mode C, as it 
exhibits the highest simulated electromechanical coupling in the coplanar geometry. 
The bilinear coupling between the qubit and the resonator is given by
\begin{align}
g_\textrm{pe} = \frac{1}{2}C_0 \sqrt{\frac{\Omega_\textrm{ge}\Omega_\textrm{m}}{C_\Sigma C}},
\end{align}
where $C_\Sigma$ is the qubit shunt capacitance, and the first qubit transition frequency is $\Omega_\textrm{ge}$ (see Supplementary Information for details). 
For optimal coupling,
% of the qubit and the mechanical resonator, 
we assume that the qubit is tuned into resonance with the mechanical oscillator, i.e., $\Omega_\textrm{ge} = \Omega_\textrm{m}$. 
The estimated coupling rate is shown in \fref{sim_fig3}(e) for various gap sizes between the electrodes and the optomechanical crystal cavity as a function of the ratio of Josephson energy $E_J$ to charging energy of the qubit $E_C$.
%with the corresponding shunt capacitance of the qubit $C_\Sigma$. 
For the qubit to exhibit low charge dispersion, a ratio $E_J/E_C\gg 1$ is required \cite{koch2007charge}. For a gap to the electrodes of \SI{50}{nm}, which can be realistically fabricated, we calculate a coupling rate of $g_\textrm{pe}/2\pi=\SI{204}{kHz}$ for $E_J/E_C = 30$, well within the transmon regime, with $C_\Sigma = \SI{91}{fF}$. 
Previously, mechanical quality factors 
%in similar GaP photonic crystal cavities
 of $Q_\textrm{m}\approx 2\times 10^5$ have been reported for similar GaP optomechanical crystal cavities 
 at the cryogenic temperature required for single phonon manipulation 
 \cite{stockill2019gallium}, which corresponds at $\Omega_\textrm{m}/2\pi = \SI{3}{GHz}$ to a mechanical damping rate of $\Gamma_\textrm{m}/2\pi = \SI{15}{kHz}$. 
We therefore estimate that the system presented here would be deep enough in the strong coupling regime to permit a faithful swap of the qubit and mechanical resonator states, if the qubit lifetime $T_1 \gtrsim \SI{10}{\micro s}$. 
Given the small contribution of the piezoelectric portion ($C_0 \approx \SI{0.79}{fF}$) to the total capacitance of such a system, we expect this $T_1$ to be achievable with current technology. 
\section*{Conclusion}
In summary, we have presented a novel platform for microwave-to-optical conversion based on single-crystal GaP optomechanical devices integrated directly on pre-fabricated niobium-on-silicon electrode structures. 
The extended mechanical breathing modes of our optomechanical crystal cavities were designed for both high optomechanical coupling rates (up to $g_0/2\pi = \SI{300}{kHz}$) and piezoelectric actuation by the microwave electrodes. 
Despite the asymmetric cavity structure and the close proximity of  the metal electrodes, high optical quality factors ($Q_0 \sim 10^5$) were obtained, placing the system in the resolved-sideband regime ($\Omega_\textrm{m}/\kappa \sim 1.1$). 
Coherent microwave-to-optical transduction in a transmission-line coupled device was demonstrated. 
The electromechanical coupling rates observed experimentally were substantially smaller than expected from simulations. The discrepancy could be fabrication-related or may be due to insufficient knowledge of the piezoelectric properties of GaP. 
In future experiments, we aim to integrate the cavity design presented here with superconducting transmon qubits, for which our simulations  indicate that strong coupling can be achieved ($g_\textrm{pe}/2\pi\approx \SI{200}{kHz}$). 
Due to the small participation ratio of the device in the overall qubit energy, we expect long qubit lifetimes.

\section*{Acknowledgements}
We gratefully acknowledge Katharina Schmeing, Yannick Baumgartner, Lukas Czornomaz, Ute Drechsler, Diana Davila and Antonis Olziersky for their valuable contributions to development of the GaP-on-insulator platform. 
We also thank  Darius Urbonas, Matthias Mergenthaler, Thomas Karg and Charles M\"{o}hl for useful discussios.  
This work was supported by the European Union's Horizon 2020 Program for Research and Innovation under grant agreement No. 722923 (Marie Curie H2020-ETN OMT) and No. 732894 (FET Proactive HOT). 
All samples were fabricated at the Binnig and Rohrer Nanotechnology Center (BRNC) at IBM Research Europe, Zurich.

\section*{Contributions}
S.H. designed the device and carried out the epitaxial growth of GaP as well as the microfabrication processes. S.H. performed FEM simulations of the photonic crystal cavities with support from A.B.. Wafer bonding of GaP was done by D.C..  Y.P. carried out chemical mechanical polishing with support from S.H.. P.S. conceived and oversaw the project. 

%\pagebreak
\bibliographystyle{apsrev4-2}
\bibliography{ref1}

\end{document}